# Reconstructed Momentum Density and Fermi Surface in $Cu_{0.9}Al_{0.1}$

M. Samsel-Czekała[1], G. Kontrym-Sznajd[1], G. Döring[2], W. Schülke[2],
S. Kaprzyk[3,4], A. Bansil[4], J. Kwiatkowska[5] and F. Maniawski[5]

[1] Institute of Low Temperature and Structure Research, PAS, 50-950 Wrocław 2, Poland

[2] Institute of Physics, University of Dortmund, D-44221 Dortmund, Germany

[3] Faculty of Physics and Nuclear Techniques, AGH, al. Mickiewicza 30, 30-073 Kraków, Poland

[4] Department of Physics, Northeastern University, Boston, Massachusetts 02115, USA

[5] Institute of Nuclear Physics, Radzikowskiego 152, 31-342 Kraków, Poland

**Abstract.** A reconstruction technique based on Radon transforms is used to obtain 3D electron momentum density $\rho(\mathbf{p})$ using nine recently measured high-resolution Compton profiles (CPs) from a $Cu_{0.9}Al_{0.1}$ disordered alloy single crystal. The method was also applied to nine corresponding theoretical CPs computed within the KKR-CPA first-principles scheme in order to show that our reconstruction procedure reproduces $\rho(\mathbf{p})$ reasonably. We comment briefly on how well a map of the Fermi surface (FS) can be obtained by folding the reconstructed $\rho(\mathbf{p})$ into the first Brillouin zone.

## Introduction

Electron momentum density, $\rho(\mathbf{p})$, can be probed via a Compton scattering experiment in which the measured Compton profile (CP), $J(p_z)$, provides the planar projection of $\rho(\mathbf{p})$ along the direction of the scattering vector $p_z$:

$$J(p_z) = \int_{-\infty}^{\infty} dp_x dp_y \, \rho(\mathbf{p}). \tag{1}$$

The 3D momentum density $\rho(\mathbf{p})$ may then be "reconstructed" by measuring CPs along various directions $p_z$. The recent possibility of obtaining high resolution CPs in wide classes of materials has rejuvenated interest in reconstruction techniques as a way of getting a handle on FS signatures and electron correlation effects in the underlying momentum density [1]. With this motivation, we have carried out extensive CP measurements on a $Cu_{0.9}Al_{0.1}$ alloy (fcc solid solution phase), and analysed the results in terms of parallel KKR-CPA computations [2,3]. Furthermore, we have reconstructed the 3D momentum density in $Cu_{0.9}Al_{0.1}$ using a technique we have recently proposed [4]. Within space limitations, this paper provides highlights of our reconstructed $\rho(\mathbf{p})$ in $Cu_{0.9}Al_{0.1}$; a detailed discussion of our theoretical and experimental study in $Cu_{0.9}Al_{0.1}$ will be taken up elsewhere.

Our reconstruction technique is based on the inversion of the Radon transforms of various quantities in terms of spherical harmonics and Jacobi polynomials [4]. $\rho(\mathbf{p})$ and $J(p_z)$ are first expanded into lattice harmonics $F_{l\nu}(\Theta,\varphi)$:

$$\rho(\mathbf{p}) = \sum_{l\nu} \rho_{l\nu}(p) F_{l\nu}(\Theta,\varphi), \tag{2}$$

$$J(p_z) \equiv J_{\beta,\alpha}(p) = \sum_{l\nu} g_{l\nu}(p) F_{l\nu}(\beta,\alpha), \tag{3}$$

where index $\nu$ distinguishes harmonics of the same order $l$ and $(\beta,\alpha)$ describe the polar and azi-muthal angles of the $p_z$-axis with respect to the reciprocal lattice. The radial components of the measured spectra are now expanded in terms of the orthogonal Jacobi polynomials $P_k^{(a,b)}$:

$$g_{lv}(p) = \sum_{k=0}^{\infty} a_{lvk}(1-p^2) P_{l+2k}^{(1,1)}(p). \tag{4}$$

The radial parts of the momentum density are then given by

$$\rho_{lv}(p) = \frac{1}{\pi} \sum_{k=0}^{\infty} a_{lvk}(1+2k+l) p^l P_k^{(0,l+1/2)}(2p^2-1). \tag{5}$$

In order to visualize the shape of the FS we have folded $\rho(\mathbf{p})$ into the reduced momentum space (i.e. the LCW-folding [5]) to obtain $\rho(\mathbf{k})$ as

$$\rho(\mathbf{k}) = \sum_{\mathbf{G}} \rho(\mathbf{p} = \mathbf{k} + \mathbf{G}) = \sum_{l} n_l(k), \tag{6}$$

where $\mathbf{k}$ denotes vectors in the first Brillouin zone and the summation may be viewed as being either over the reciprocal lattice vectors $\mathbf{G}$, or over the band index $l$. The occupation number $n$ is unity for filled states and zero otherwise.

**Results**

CPs were measured at the Compton spectrometer of the ESRF (Grenoble) along 9 directions given by angles $(\Theta, \varphi)$: $(90^0, 0^0) \equiv [100]$; $(90^0, 10^0)$; $(90^0, 20^0)$; $(90^0, 45^0) \equiv [110]$; $(80^0, 45^0)$; $(54.74^0, 45^0) \equiv [111]$; $(63.53^0, 39.43^0)$; $(72.50^0, 34.67^0)$; $(81.58^0, 30.36^0)$. The corresponding theoretical CPs were computed within the fully selfconsistent KKR-CPA framework [2,3] and are highly accurate; we avoid further experimental and theoretical details here in the interest of brevity. Turning to the question of reconstruction, we emphasize that it is only for special sets of directions that one can use as many terms as the number of profiles in the expansions of Eqs. 2 and 3; in general the number of terms is smaller (e.g. [6]), and in the present case we had to restrict these expansions to the first 6 lattice harmonics. Fig. 1 provides illustrative results. As seen from the right side of the figure, momentum densities reconstructed from 9 theoretical CPs along 3 different directions are quite similar to the $\rho(\mathbf{p})$ computed directly. Reconstructed momentum densities based on either the theoretical or experimental CPs correctly display salient features such as the lack of a Fermi break along [111] due to the presence of the neck, the anisotropy between the [100] and [110] and the prominent Umklapp components along [100], some obvious discrepancies notwithstanding. These results indicate the overall reasonableness of our reconstruction procedure.

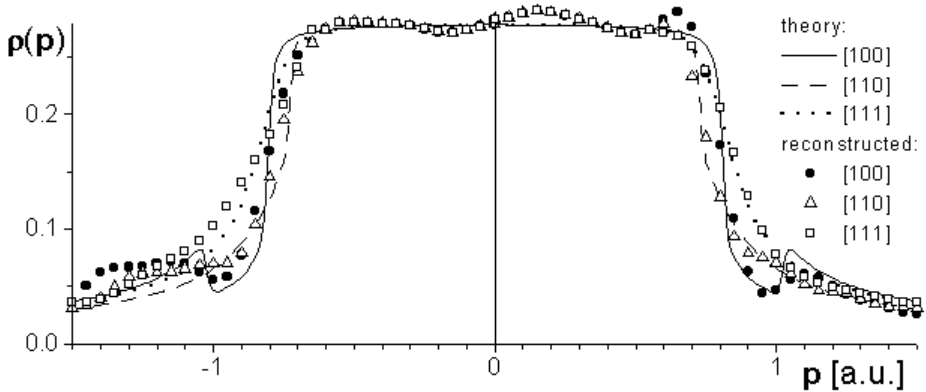

Fig. 1 Theoretically computed KKR-CPA momentum density $\rho(\mathbf{p})$ in $Cu_{0.9}Al_{0.1}$ along three high symmetry directions (various lines indicated in the legend on *both* the right and left hand sides) is compared with the corresponding reconstructed results based on 9 computed CPs (right side) and 9 experimental CPs (left side).

We consider next the LCW-folded momentum density $\rho(\mathbf{k})$ (see, Eq. 6) based on $\rho(\mathbf{p})$ reconstructed from theoretical and experimental CPs. Results in the $(1\bar{1}0)$ plane are presented in Fig. 2 where the summation in Eq. 6 has been carried out over a cube in momentum space of sides 5.0 a.u.. The folded $\rho(\mathbf{k})$ derived from the theoretical CPs (Fig. 2a) as well as those employing the experimental CPs (Fig. 2b) reasonably display signatures of the well known FS of Cu. The prominent peak seen at the zone center in both Figs. 2a and 2b seems to partly reflect the slower convergence of the summation in Eq. 6 for states of d character which possess a fairly long range in momentum space due to their localized nature [7]. This point however bears further investigation. A similar effect may also be at play in creating additional peaks at the centers of the necks at the L-points in Fig. 2a. In the case of $\rho(\mathbf{k})$ obtained from the experimental CPs, Fig. 2b, we see fluctuations of density throughout the Brillouin zone; these are the result of a relatively high experimental error in the reconstructed momentum density.

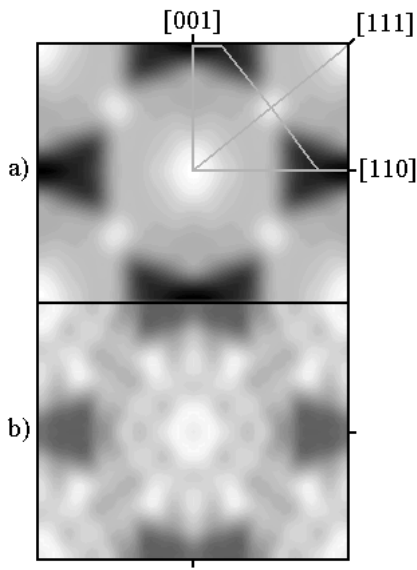

Fig. 2 Folded momentum density $\rho(\mathbf{k})$ in $Cu_{0.9}Al_{0.1}$ in the $(1\bar{1}0)$ plane obtained from 9 theoretical (top part) and 9 experimental (bottom part) CPs. Lighter colors denote higher values.

## Conclusions

We have applied a new reconstruction procedure based on the use of Radon transforms to high resolution CPs measured from a disordered $Cu_{0.9}Al_{0.1}$ alloy single crystal. The 3D momentum density $\rho(\mathbf{p})$ has been reconstructed from 9 experimental profiles as well as from the 9 corresponding theoretical profiles computed within the framework of the fully selfconsistent KKR-CPA scheme. Comparisons between the $\rho(\mathbf{p})$'s reconstructed from the experimental and theoretical profiles with each other and with the results of a direct computation (without a reconstruction) show that our procedure reasonably reproduces the salient features of the momentum density in this system. We have also applied the LCW-folding to the reconstructed $\rho(\mathbf{p})$'s in order to obtain a map of the FS. These results indicate that while the folded density yields a resonable description of the FS, convergence properties of the folding require further study.

**Acknowledgements.** It is a pleasure to acknowledge important conversations with Bernardo Barbiellini and Peter Mijnarends. This work was supported by the Polish State Committee for Scientific Research (Grant 2 P03B 083 16), the German Federal Ministry of Education and Research Contract 05 ST 8 HRA, the US Department of Energy contract W-31-109-ENG-38, a travel grant from NATO, and benefited from the allocation of supercomputer time at the NERSC and the Northeastern University Advanced Scientific Computation Center (NU-ASCC).

Corresponding author: M. Samsel-Czekała, Tel.: (48-71) 34 350 21/322, Fax: (48-71) 441 029, e-mail: samsel@int.pan.wroc.pl